# Representation of wave fields in a graded multilayer medium

M.I. Ayzatsky

National Science Center Kharkov Institute of Physics and Technology (NSC KIPT), 61108, Kharkov, Ukraine

E-mail: mykola.aizatsky@gmail.com

In this work we present a new approach in representation of wave fields in nonuniform 1-D multilayer medium. This approach is based on the use of a modified homogeneous basis. A new form of coupled equations for describing nonuniform one-directional photonic crystals is proposed. This description can be especially useful to study radiation that originates when fast electrons move in such medium

## 1. INTRODUCTION

Photonic crystals are (quasiperiodic) dielectric structures composed of alternating high and low index of refraction dielectric materials, with periodicity on the order of the wavelength of light. The one-dimensional photonic crystal represents a set of layers with different parameters that constitute a periodic structure.

Of particular interest are studies of the characteristics of inhomogeneous photonic crystals (see, for example, [1,2,3,4,5,6,7,8,9,10,11,12]). The most used method for describing inhomogeneous 1-D photonic crystals is the matrix method [3]. The coupled-mode theory has also been proposed to study slow taper transitions in photonic crystals [6,9,12]. In this work we propose a continuous coupled-mode theory that can be used to study 1-D inhomogeneous photonic crystals with any inhomogeneities. This description can be especially useful to study radiation that originates when fast electrons move in such medium (see, for example [13,14]). A similar approach has recently been proposed to describe structured waveguides [15].

## 2. FIELD EXPANSION

We consider a graded 1-D multilayer media that consists of parallel, linear, homogeneous, and isotropic layers. The axis $z$ is chosen along the direction normal to the layer interfaces.

Let's combine layers in pairs. Each pair (we will call the pair as a combined layer) consists of the first layer with thickness $d_{1,k}$ and permittivity $\varepsilon_{1,k}$ and the second layer with thickness $d_{2,k}$ and permittivity $\varepsilon_{2,k}$. We will assume that all quantities have time and $x$ variations given by $\exp(ik_x x - i\omega t)$. Electromagnetic fields in considered medium are the solutions of the Maxwell's equations

$$rot\,\vec{E} = i\omega\mu_0 \vec{H}, \tag{1}$$

$$rot\,\vec{H} = -i\omega\varepsilon_0 \hat{\varepsilon}\vec{E} + \vec{j}, \tag{2}$$

where

$$\hat{\varepsilon} = \begin{cases} \varepsilon_{1,k}, & z_k < z < z_k + d_{1,k}, \\ \varepsilon_{2,k}, & z_k + d_{1,k} < z < z_{k+1}, \end{cases} \tag{3}$$

$z_k = \sum_{\infty}^{k-1} D_s$, $D_k = d_{1,k} + d_{2,k}$. For convenience, we designate $g_{1,k} = d_{1,k}$, $g_{2,k} = \varepsilon_{1,k}$, $g_{3,k} = d_{2,k}$, $g_{4,k} = \varepsilon_{2,k}$.

Consider a periodic multilayer with constant periodic permittivity in layers $\varepsilon_1$, $\varepsilon_2$ and layer thicknesses $d_1$ and $d_2$ ($D = d_1 + d_2$). In homogeneous periodic medium there are two eigen waves: forward (+) and backward (-)

$$\vec{E}^{\pm} = \exp\{\pm\gamma(g_1,...,g_4)z\} \begin{cases} \vec{E}_1^{\pm}(z,g_1,...,g_4), & kD < z < kD + d_1, \\ \vec{E}_2^{\pm}(z,g_1,...,g_4), & kD + d_1 < z < (k+1)kD, \end{cases}$$

$$\vec{H}^{\pm} = \exp\{\pm\gamma(g_1,...,g_4)z\} \begin{cases} \vec{H}_1^{\pm}(z,g_1,...,g_4), & kD < z < kD + d_1, \\ \vec{H}_2^{\pm}(z,g_1,...,g_4), & kD + d_1 < z < (k+1)kD, \end{cases} \tag{4}$$

where $g_1 = d_1$, $g_2 = \varepsilon_1$, $g_3 = d_2$, $g_4 = \varepsilon_2$.

$\vec{E}_q^{\pm}, \vec{H}_q^{\pm}$ ($q = 1,2$) are the periodic functions of $z$ and we can consider that they are the functions of $\tilde{z} = z - kD$. $\vec{E}_q^{\pm}, \vec{H}_q^{\pm}$ are the solutions of such equations

$$\begin{aligned} rot\,\vec{H}_q^{\pm} \pm \gamma\left[\vec{e}_z \vec{H}_q^{\pm}\right] &= -i\omega\varepsilon_0 \varepsilon_q \vec{E}_q^{\pm}, \\ rot\,\vec{E}_q^{\pm} \pm \gamma\left[\vec{e}_z \vec{E}_q^{\pm}\right] &= i\omega\mu_0 \vec{H}_q^{\pm}, \end{aligned} \tag{5}$$

where $0 < \tilde{z} < d_1$ for $q = 1$ and $d_1 < \tilde{z} < d_1 + d_2 = D$ for $q = 2$.

$(\vec{E}^{\pm}, \vec{H}^{\pm})$ satisfy certain boundary conditions at the interfaces $z = kD$ and $z = kD + d_1$

For each fixed $z$ there are a set of $g_i^{(l)}$ (we will call them local parameters), that determine the layer characteristics. The remaining parameters we will call global $g_i^{(g)}$. For example, for $kD < z < kD + d_1$ the parameters $g_1 = d_1$ and $g_2 = \varepsilon_1$ are the local, the parameters $g_3 = d_2$ and $g_4 = \varepsilon_2$ are the global ones. The division of the parameters $g_i$ into local and global ones depends on $z$.

We can introduce new vector functions $\vec{E}_q^{\pm(z)} = \vec{E}_q^{\pm}\left(\tilde{z}, g_1^{(z)}(z), \ldots, g_4^{(z)}(z)\right)$, $\vec{H}_q^{\pm(z)} = \vec{H}_q^{\pm}\left(\tilde{z}, g_1^{(z)}(z), \ldots, g_4^{(z)}(z)\right)$, ($q = 1, 2$) by formally replacing the constants $g_i$ in (4) with functions $g_i^{(z)}(z)$, where $g_i^{(z)}(z)$ and its derivatives are continuous functions of $z$. We are also introducing a new function $\gamma^{(z)} = \gamma\left(g_1^{(z)}(z), \ldots, g_4^{(z)}(z)\right)$.

The vector functions $\vec{E}_q^{\pm(z)}$, $\vec{H}_q^{\pm(z)}$ are no longer the solutions to Maxwell equations. Indeed, as

$$\frac{\partial \vec{E}_q^{\pm(z)}}{\partial z} = \frac{\partial \vec{E}_{q,g}^{\pm(z)}}{\partial z} + \sum_i \frac{\partial \vec{E}_q^{\pm(z)}}{\partial g_i} \frac{dg_i}{dz}, \tag{6}$$

where $\vec{E}_{q,g}^{\pm(z)} = \vec{E}_q^{\pm}\left(\tilde{z}, g_i = const\right)$, then

$$rot\, \vec{E}_q^{\pm(z)} = i\omega\mu_0 \vec{H}_q^{\pm(z)} - \gamma^{(z)}\left[\vec{e}_z \vec{E}_q^{\pm(z)}\right] + \vec{E}_q^{\pm(\nabla)}, \tag{7}$$

$$rot\, \vec{H}_q^{\pm(z)} = -i\omega\varepsilon_0 \varepsilon_q^{(z)} \vec{E}_q^{\pm(z)} - \gamma^{(z)}\left[\vec{e}_z \vec{H}_q^{\pm(z)}\right] + \vec{H}_q^{\pm(\nabla)}, \tag{8}$$

where

$$\vec{E}_q^{\pm(\nabla)} = \sum_i \left[\vec{e}_z \frac{\partial \vec{E}_q^{\pm(z)}}{\partial g_i}\right] \frac{dg_i^{(z)}}{dz},$$

$$\vec{H}_q^{\pm(\nabla)} = \sum_i \left[\vec{e}_z \frac{\partial \vec{H}_q^{\pm(z)}}{\partial g_i}\right] \frac{dg_i^{(z)}}{dz}. \tag{9}$$

Before using the vector functions $\vec{E}_q^{\pm(z)}$, $\vec{H}_q^{\pm(z)}$ to describe the non-periodic layered medium (3), it is necessary to define a procedure for choosing functions $g_i^{(z)}(z)$. As the parameters $g_{1,k}$ and $g_{2,k}$ describe the properties of the first layer in the combined layer $k$, the functions $g_1^{(z)}(z)$ and $g_2^{(z)}(z)$ on the interval $\left(z_k, z_k + d_{1,k}\right)$ must be chosen constant: $g_1^{(z)}(z) = g_{1,k}$, $g_2^{(z)}(z) = g_{2,k}$. By analogy, the functions $g_3^{(z)}(z)$ and $g_4^{(z)}(z)$ on the interval $\left(z_k + d_{1,k}, z_{k+1}\right)$ must also be chosen constant: $g_3^{(z)}(z) = g_{2,k}$, $g_4^{(z)}(z) = g_{4,k}$.

In the first layer (on the interval $\left(z_k, z_k + d_{1,k}\right)$) the functions $g_3^{(z)}(z)$ and $g_4^{(z)}(z)$ can be chosen arbitrarily. But as these functions must to be continues (with derivatives) and on the neighbor intervals they are constant (on the left $g_3^{(z)}(z) = g_{3,k-1}$, $g_4^{(z)}(z) = g_{4,k-1}$; on the right $g_3^{(z)}(z) = g_{3,k}$, $g_4^{(z)}(z) = g_{4,k}$) we need to connect two horizontal lines with a function that has as many zero derivatives as possible at both ends of the interval. If we want to have $N$ zero derivatives, we can use a polynomial of degree ($2N+1$). By analogy, we must choose $g_1^{(z)}(z)$ and $g_2^{(z)}(z)$ on the interval $\left(z_k + d_{1,k}, z_{k+1}\right)$. We can use, for example, functions like this ($N = 2$)

$$g_1^{(z)}(z) = \begin{cases} g_{1,k}, & z_k < z < z_k + d_{1,k}, \\ g_{1,k} + \left(g_{1,k+1} - g_{1,k}\right)\left\{10\frac{\tilde{z}^3}{d_{2,k}^3} - 15\frac{\tilde{z}^4}{d_{2,k}^4} + 6\frac{\tilde{z}^5}{d_{2,k}^5}\right\}, & \tilde{z} = z - z_k, \; z_{k,1} + d_{1,k} < z < z_{k+1}, \end{cases} \tag{10}$$

$$g_3^{(z)}(z) = \begin{cases} g_{3,k-1} + \left(g_{3,k} - g_{3,k-1}\right)\left\{10\frac{\tilde{z}^3}{d_{1,k}^3} - 15\frac{\tilde{z}^4}{d_{1,k}^4} + 6\frac{\tilde{z}^5}{d_{1,k}^5}\right\}, & \tilde{z} = z - z_k, \; z_k < z < z_k + d_{1,k}, \\ g_{3,k}, & z_k + d_{1,k} < z < z_{k+1}, \end{cases} \tag{11}$$

Since $\vec{H}_q^{\pm(z)}$ are continuous everywhere, therefore, we can represent the solution $\vec{H}$ of equations (1) and (2) as

$$\vec{H} = C^+(z)\begin{cases} \vec{H}_1^{+(z)}, & z_k < z < z_k + d_{1,k} \\ \vec{H}_2^{+(z)}, & z_k + d_{1,k} < z < z_{k+1} \end{cases} + C^-(z)\begin{cases} \vec{H}_1^{-(z)}, & z_k < z < z_k + d_{1,k} \\ \vec{H}_2^{-(z)}, & z_k + d_{1,k} < z < z_{k+1} \end{cases}. \tag{12}$$

Using (8), we obtain

$$rot\, \vec{H} = \begin{cases} rot\, \vec{H}_1 & z_k < z < z_k + d_{1,k}, \\ rot\, \vec{H}_2, & z_k + d_{1,k} < z < z_{k+1}, \end{cases} \tag{13}$$

where

$$rot\,\vec{H}_q = -i\omega\varepsilon_0\varepsilon_{q,k}\vec{E}_q^{+(z)}C^+ - i\omega\varepsilon_0\varepsilon_{q,k}\vec{E}_q^{-(z)}C^- + \frac{dC^+}{dz}\left[\vec{e}_z\vec{H}_q^{+(z)}\right] + \frac{dC_{-s}}{dz}\left[\vec{e}_z\vec{H}_q^{-(z)}\right] - $$
$$\gamma^{(z)}\left[\vec{e}_z\vec{H}_q^{+(z)}\right]C^+ + \gamma^{(z)}\left[\vec{e}_z\vec{H}_q^{-(z)}\right]C^- + \vec{H}_q^{+(\nabla)}C^+ + \vec{H}_q^{-(\nabla)}C_s \qquad (14)$$

Then from (2) we find

$$\vec{E} = \begin{cases} \vec{E}_1, & z_k < z < z_k + d_{1,k}, \\ \vec{E}_2, & z_k + d_{1,k} < z < z_{k+1}, \end{cases} \qquad (15)$$

where

$$\vec{E}_q = C^+\vec{E}_q^{+(z)} + C^-\vec{E}_q^{-(z)} - $$
$$\frac{1}{i\omega\varepsilon_0\varepsilon_{q,k}}\left\{\left(\frac{dC^+}{dz} - \gamma^{(z)}C^+\right)\left[\vec{e}_z\vec{H}_q^{+(z)}\right] + \left(\frac{dC^-}{dz} + \gamma^{(z)}C^-\right)\left[\vec{e}_z\vec{H}_q^{-(z)}\right] + \vec{H}_q^{+(\nabla)}C^+ + \vec{H}_q^{-(\nabla)}C^-\right\} + \frac{\vec{j}}{i\omega\varepsilon_0\varepsilon_{q,k}} \qquad (16)$$

Introducing two new functions $C^+(z)$ and $C^-(z)$ we can impose an additional condition. We choose this

$$\left\{\left(\frac{dC^+}{dz} - \gamma^{(z)}C^+\right)\left[\vec{e}_z\vec{H}_q^{+(z)}\right] + \left(\frac{dC^-}{dz} + \gamma^{(z)}C^-\right)\left[\vec{e}_z\vec{H}_q^{-(z)}\right] + \vec{H}_q^{+(\nabla)}C^+ + \vec{H}_q^{-(\nabla)}C^-\right\} = \vec{j}_\perp. \qquad (17)$$

Then (15) takes the form

$$\vec{E} = C^+\vec{E}_q^{+(z)} + C^-\vec{E}_q^{-(z)} + \frac{\vec{j}_z}{i\omega\varepsilon_0\varepsilon_{q,k}} \qquad (18)$$

We remind that $z_k < z < z_k + d_{1,k}$ for $q = 1$ and $z_k + d_{1,k} < z < z_{k+1}$ for $q = 2$.

Therefore, we obtained such representations

$$\vec{E} = C^+\vec{E}_q^{+(z)} + C^-\vec{E}_q^{-(z)} + \frac{\vec{j}_z}{i\omega\varepsilon_0\varepsilon_{q,k}},$$
$$\vec{H} = C^+\vec{H}_q^{+(z)} + C^-\vec{H}_q^{-(z)}. \qquad (19)$$

They will be unambiguous if $C^+$ and $C^-$ can be found in a unique way from these representations.

Let's multiply the first equation by $\vec{H}_q^{+(z)}$, the second by $\vec{E}_q^{-(z)}$ and add the results

$$\left[\vec{\tilde{E}}\vec{H}_q^{+(z)}\right] + \left[\vec{H}\vec{E}_q^{+(z)}\right] = C^+\left[\vec{E}_q^{+(z)}\vec{H}_q^{+(z)}\right] + C^+\left[\vec{H}_q^{+(z)}\vec{E}_q^{+(z)}\right] + C^-\left[\vec{E}_q^{-(z)}\vec{H}_q^{+(z)}\right] + C^-\left[\vec{H}_q^{-(z)}\vec{E}_q^{+(z)}\right], \qquad (20)$$

where $\vec{\tilde{E}} = \vec{E} - \dfrac{\vec{j}_z}{i\omega\varepsilon_0\varepsilon_{q,k}}$

The vector $\left(\left[\vec{E}_q^{-(z)}\vec{H}_q^{+(z)}\right] + \left[\vec{H}_q^{-(z)}\vec{E}_q^{+(z)}\right]\right)$ can be represented as $\left|\left[\vec{E}_q^{-(z)}\vec{H}_q^{+(z)}\right] + \left[\vec{H}_q^{-(z)}\vec{E}_q^{+(z)}\right]\right|\vec{u}_q^{(z)}$, where $|\,|$ is the module and $\vec{u}_q^{(z)}$ is the unit vector. Then from equation (20) we obtain

$$C^- = \left(\vec{u}^{(z)}\left[\vec{\tilde{E}}\vec{H}_q^{+(z)}\right] + \vec{u}^{(z)}\left[\vec{H}\vec{E}_q^{+(z)}\right]\right)\left|\left[\vec{E}_q^{-(z)}\vec{H}_q^{+(z)}\right] + \left[\vec{H}_q^{-(z)}\vec{E}_q^{+(z)}\right]\right|^{-1}. \qquad (21)$$

Therefore, if $\left|\left[\vec{E}_q^{-(z)}\vec{H}_q^{+(z)}\right] + \left[\vec{H}_q^{-(z)}\vec{E}_q^{+(z)}\right]\right| \neq 0$, there is only one way to find $C^+$ ($C^-$) and the relations (19) give the correct representation.

Insertion (19) into the equation (1) gives

$$rot\,\vec{E} = rot\left(C^+\vec{E}_q^{+(z)} + C^-\vec{E}_q^{-(z)}\right) = \left\{\left(\frac{dC^+}{dz} - \gamma^{(z)}C^+\right)\left[\vec{e}_z\vec{E}_q^{+(z)}\right] + \left(\frac{dC^-}{dz} + \gamma^{(z)}C^-\right)\left[\vec{e}_z\vec{E}_q^{-(z)}\right] + C^+\vec{E}_q^{+(\nabla)} + C^-\vec{E}_q^{-(\nabla)}\right\} + $$
$$i\omega\mu_0\left(C^+\vec{H}_q^{+(z)} + C^-\vec{H}_q^{-(z)}\right) + \frac{1}{i\omega\varepsilon_0\varepsilon_{q,k}}rot\,\vec{j}_z = i\omega\mu_0\left(C^+\vec{H}_q^{+(z)} + C^-\vec{H}_q^{-(z)}\right). \qquad (22)$$

From this equality it follows

$$\left(\frac{dC^+}{dz} - \gamma^{(z)}C^+\right)\left[\vec{e}_z\vec{E}_q^{+(z)}\right] + \left(\frac{dC^-}{dz} + \gamma^{(z)}C^-\right)\left[\vec{e}_z\vec{E}_q^{-(z)}\right] + C^+\vec{E}_q^{+(\nabla)} + C^-\vec{E}_q^{-(\nabla)} = -\frac{1}{i\omega\varepsilon_0\varepsilon_{q,k}}rot\,\vec{j}_z. \qquad (23)$$

Multiplying the equation (17) by $\vec{E}_q^{-(z)}$ ($\vec{E}_q^{+(z)}$), the equation (23) by $\vec{H}_q^{-(z)}$ ($\vec{H}_q^{+(z)}$) and adding the results, we get

$$N_q^{(z)}\left(\frac{dC^+}{dz}-\gamma^{(z)}C^+\right)+C^+\left(\vec{E}_q^{+(\nabla)}\vec{H}_q^{-(z)}+\vec{H}_q^{+(\nabla)}\vec{E}_q^{-(z)}\right)+C^-\left(\vec{E}_q^{-(\nabla)}\vec{H}_q^{-(z)}+\vec{H}_q^{-(\nabla)}\vec{E}_q^{-(z)}\right)=\vec{E}_q^{-(z)}\vec{j}_\perp-\frac{1}{i\omega\varepsilon_0\varepsilon_{q,k}}\vec{H}_q^{-(z)}rot\,\vec{j}_z,$$

$$-N_q^{(z)}\left(\frac{dC^-}{dz}-\gamma^{(z)}C^-\right)+C^+\left(\vec{E}_q^{+(\nabla)}\vec{H}_q^{+(z)}+\vec{H}_q^{+(\nabla)}\vec{E}_q^{+(z)}\right)+C^-\left(\vec{E}_q^{-(\nabla)}\vec{H}_q^{+(z)}+\vec{H}_q^{-(\nabla)}\vec{E}_q^{+(z)}\right)=\vec{E}_q^{+(z)}\vec{j}_\perp-\frac{1}{i\omega\varepsilon_0\varepsilon_{q,k}}\vec{H}_q^{+(z)}rot\,\vec{j}_z,$$

(24)

where $N_q^{(z)}=\vec{e}_z\left(\left[\vec{E}_q^{+(z)}\vec{H}_q^{-(z)}\right]+\left[\vec{H}_q^{+(z)}\vec{E}_q^{-(z)}\right]\right)$.

It can be shown that

$$\vec{H}_q^{\pm(z)}rot\,\vec{j}_z=-i\omega\varepsilon_0\varepsilon_{q,k}\vec{E}_q^{\pm(z)}\vec{j}_z. \quad (25)$$

Therefore, the final equations have such form

$$N_q^{(z)}\left(\frac{dC^+}{dz}-\gamma^{(z)}C^+\right)+C^+\left(\vec{E}_q^{+(\nabla)}\vec{H}_q^{-(z)}+\vec{H}_q^{+(\nabla)}\vec{E}_q^{-(z)}\right)+C^-\left(\vec{E}_q^{-(\nabla)}\vec{H}_q^{-(z)}+\vec{H}_q^{-(\nabla)}\vec{E}_q^{-(z)}\right)=\vec{E}_q^{-(z)}\vec{j},$$

$$-N_q^{(z)}\left(\frac{dC^-}{dz}-\gamma^{(z)}C^-\right)+C^+\left(\vec{E}_q^{+(\nabla)}\vec{H}_q^{+(z)}+\vec{H}_q^{+(\nabla)}\vec{E}_q^{+(z)}\right)+C^-\left(\vec{E}_q^{-(\nabla)}\vec{H}_q^{+(z)}+\vec{H}_q^{-(\nabla)}\vec{E}_q^{+(z)}\right)=\vec{E}_q^{+(z)}\vec{j}.$$

(26)

For TE waves ($p$-polarization, $E_y, H_x, H_z$) inside the propagation band ($\gamma=ih$) these equations ($\vec{j}=0$) can be transformed to the following

$$N_{q,p}^{(z)}\left\{\frac{dC^+}{dz}-\left(\gamma^{(z)}+\tilde{z}\frac{d\gamma^{(z)}}{dz}\right)C^+ + \frac{1}{2N_{q,p}^{(z)}}\frac{dN_q^{(z)}}{dz}C^+\right\}+C^+W_q^{+(z)}+C^-\Lambda_q^{-(z)}=0,$$

$$N_{q,p}^{(z)}\left\{\frac{dC^-}{dz}+\left(\gamma^{(z)}+\tilde{z}\frac{d\gamma^{(z)}}{dz}\right)C^- + \frac{1}{N_{q,p}^{(z)}2}\frac{dN_q^{(z)}}{dz}C^-\right\}-C^-W_q^{-(z)}-C^+\Lambda_q^{+(z)}=0,$$

(27)

where $\tilde{z}=z-z_k$,

$$W_q^{\pm(z)}=\frac{\sigma_{q,p}^{(z)}}{Z_0}2i\sum_i\frac{dg_i^{(z)}}{dz}\text{Im}\left\{\left(G_{q,y}^{+,-(z)}\right)^*\frac{\partial G_{q,y}^{+,-(z)}}{\partial g_i^{(z)}}-\left(G_{q,y}^{+,+(z)}\right)^*\frac{\partial G_{q,y}^{+,+(z)}}{\partial g_i^{(z)}}\right\} \quad (28)$$

$$\Lambda_q^{\pm(z)}=\frac{\sigma_{q,p}^{(z)}}{Z_0}2\exp\{\mp 2\gamma^{(z)}\tilde{z}\}\sum_i\frac{dg_i^{(z)}}{dz}\left(G_{q,y}^{\pm,+(z)}\frac{\partial G_{q,y}^{\pm,-(z)}}{\partial g_i^{(z)}}-G_{q,y}^{\pm,-(z)}\frac{\partial G_{q,y}^{\pm,+(z)}}{\partial g_i^{(z)}}\right) \quad (29)$$

$$N_p^{(z)}=-2\frac{\sigma_{q,p}}{Z_0}\left(G_{q,y}^{+,+(z)}G_{q,y}^{-,-(z)}-G_{q,y}^{+,-(z)}G_{q,y}^{-,+(z)}\right) \quad (30)$$

Coefficients $G_{q,y}^{\pm,\pm}, G_{q,y}^{\pm,\mp}$ are given in the Appendix 1, where $d_1,\varepsilon_1,d_2,\varepsilon_2,$ must be changed to $d_1^{(z)}(z),\varepsilon_1^{(z)}(z),d_2^{(z)}(z),\varepsilon_2^{(z)}(z)$

Consider the case $k_x=0$ and $d_{1,k}=d_{2,k}=d\to 0$, ($k_{q,z}^{(z)}d\ll 1$). The solution to equation (42) (Appendix 1) can be obtained in analytical form

$$\rho^\pm\approx 1\pm id\sqrt{2\left(k_{1,z}^2+k_{2,z}^2\right)}. \quad (31)$$

Using the definition of $\gamma^{(z)}$, we get

$$\gamma^{(z)}=i\frac{\omega}{c}\sqrt{\frac{\varepsilon_1^{(z)}+\varepsilon_2^{(z)}}{2}}=i\frac{\omega}{c}\sqrt{\varepsilon^{(z)}}, \quad (32)$$

where

$$\varepsilon^{(z)}=\frac{\varepsilon_1^{(z)}+\varepsilon_2^{(z)}}{2} \quad (33)$$

As $\varepsilon_1^{(z)}$ is constant over $z_k<z<z_k+d$ and $\varepsilon_2^{(z)}$ is constant over $z_k+d<z<z_{k+1}$, then we have

$$\frac{d\varepsilon^{(z)}}{dz}=\begin{cases}\frac{1}{2}\frac{d\varepsilon_2^{(z)}}{dz}, & z_k<z<z_k+d,\\ \frac{1}{2}\frac{d\varepsilon_1^{(z)}}{dz}, & z_k+d<z<z_{k+1}.\end{cases} \quad (34)$$

If $\left|\varepsilon_1^{(z)}-\varepsilon_2^{(z)}\right|/\varepsilon_1^{(z)}\ll 1$, then system (27) goes into this

$$\frac{dC^+}{dz} - \gamma^{(z)}C^+ + \frac{1}{2N_q^{(z)}}\frac{dN_q^{(z)}}{dz}C^+ + C^-\tilde{\Lambda}_q^{-(z)} = 0,$$

$$\frac{dC^-}{dz} + \gamma^{(z)}C^- + \frac{1}{2N_q^{(z)}}\frac{dN_q^{(z)}}{dz}C^- - C^+\tilde{\Lambda}_q^{+(z)} = 0,$$
(35)

where

$$\tilde{\Lambda}_q^{\pm(z)} \approx \pm\frac{1}{8}\frac{1}{\varepsilon^{(z)}}\frac{d\varepsilon_q^{(z)}}{dz} \tag{36}$$

$$N_q^{(z)} \approx -2\frac{\sqrt{\varepsilon^{(z)}}}{Z_0}G_{1,y}^{+,+}G_{1,y}^{-,-}. \tag{37}$$

Using relation (34), we finally get

$$\frac{dC^+}{dz} - i\frac{\omega}{c}\sqrt{\varepsilon^{(z)}}C^+ + \frac{1}{4\varepsilon^{(z)}}\frac{d\varepsilon^{(z)}}{dz}C^+ - \frac{1}{4\varepsilon^{(z)}}\frac{d\varepsilon^{(z)}}{dz}C^- = 0,$$

$$\frac{dC^-}{dz} + i\frac{\omega}{c}\sqrt{\varepsilon^{(z)}}C^- + \frac{1}{4\varepsilon^{(z)}}\frac{d\varepsilon^{(z)}}{dz}C^- - \frac{1}{4\varepsilon^{(z)}}\frac{d\varepsilon^{(z)}}{dz}C^+ = 0,$$
(38)

This system coincides with the system which describes the nonuniform dielectric (see Appendix 2).

Therefore, only if $\left|\varepsilon_1^{(z)} - \varepsilon_2^{(z)}\right|/\varepsilon_1^{(z)} \ll 1$ in the long wavelength limit the nonuniform multilayer medium can be considered as a nonuniform artificial dielectric.

## 2. CONCLUSIONS

Generalization of the theory of coupled modes makes it possible to describe 1-D multilayer medium by continues equations. Based on a set of eigen waves of a homogeneous periodic medium, a new basis of vector functions is introduced that takes into account the non-periodicity. By representing the total field as the sum of these functions with unknown scalar coefficients, a system of coupled equations was obtained that determines the dependence of these coefficients on the longitudinal coordinate. The proposed coupled-mode theory can be used to study 1-D inhomogeneous photonic crystals with any inhomogeneities.

## APPENDIX 1

For TE waves ($p$-polarization, $E_y, H_x, H_z$) in a periodic multilayer with constant periodic permittivity in layers $\varepsilon_1, \varepsilon_2$ and layer widths $d_1, d_2$, we can write the components of eigen waves as

$$E_{q,y,k}^{\pm,\Sigma} = \exp(\pm\gamma z)E_{q,y,k}^{\pm},$$
$$H_{q,x,k}^{\pm,\Sigma} = \exp(\pm\gamma z)H_{q,x,k}^{\pm},$$
(39)

where

$$E_{q,y,k}^{\pm} = \left\{G_{q,y}^{\pm,+}\exp\{(\mp\gamma + ik_{q,z})\tilde{z}\} + G_{q,y}^{\pm,-}\exp\{(\mp\gamma - ik_{q,z})\tilde{z}\}\right\},$$

$$H_{q,x,k}^{\pm} = -\frac{\sigma_{q,p}}{Z_0}\left\{G_{q,y}^{\pm,+}\exp\{(\mp\gamma + ik_{q,z})\tilde{z}\} - G_{q,y}^{\pm,-}\exp\{(\mp\gamma - ik_{q,z})\tilde{z}\}\right\}.$$
(40)

Expressions (39) for eigen waves we can be rewritten as

$$E_{q,y,k}^{\pm,\Sigma} = \exp(\pm\gamma z)\left(G_{q,y}^{\pm,+}\exp\{(\mp\gamma + ik_{q,z})\tilde{z}\} + G_{q,y}^{\pm,-}\exp\{(\mp\gamma - ik_{q,z})\tilde{z}\}\right) =$$
$$= \left(\rho^{\pm}\right)^k\left\{G_{q,y}^{\pm,+}\exp(ik_{q,z}\tilde{z}) + G_{q,y}^{\pm,-}\exp(-ik_{q,z}\tilde{z})\right\},$$

$$H_{q,x,k}^{\pm,\Sigma} = -\exp(\pm\gamma z)\frac{\sigma_{q,p}}{Z_0}\left\{G_{q,y}^{\pm,+}\exp\{(\mp\gamma + ik_{q,z})\tilde{z}\} - G_{q,y}^{\pm,-}\exp\{(\mp\gamma - ik_{q,z})\tilde{z}\}\right\}$$
$$= -\left(\rho^{\pm}\right)^k\frac{\sigma_{q,p}}{Z_0}\left\{G_{q,y}^{\pm,+}\exp(ik_{q,z}\tilde{z}) - G_{q,y}^{\pm,-}\exp(-ik_{q,z}\tilde{z})\right\},$$
(41)

where $k_{q,z}^2 = \left(\frac{\omega^2}{c^2}\varepsilon_q - k_x^2\right)$, $\sigma_{q,p} = \sqrt{\varepsilon_q - \frac{c^2 k_x^2}{\omega^2}}$, $\tilde{z} = z - kD$, $0 < \tilde{z} < d_1$ for $q = 1$ and $d_1 < \tilde{z} < d_1 + d_2 = D$ for $q = 2$.

The Floquet constant ($\rho = \exp(\gamma D)$) is determined by such equation

$$\left(\rho^{\pm}\right)^2 - \rho^{\pm}2Q + 1 = 0, \tag{42}$$

where $Q = \left\{\alpha_p^{(+)2}\cos(k_{2,z}d_2 + k_{1,z}d_1) - \alpha_p^{(-)2}\cos(-k_{2,z}d_2 + k_{1,z}d_1)\right\}\left(\alpha_p^{(+)2} - \alpha_p^{(-)2}\right)^{-1}$, $\alpha_p^{(\pm)} = 1 \pm \dfrac{\sigma_{1,p}}{\sigma_{2,p}}$.

If we choose as independent amplitude $G_{1,y}^{+,+}$ for the forward eigen wave and $G_{1,y}^{-,-} = \left(G_{1,y}^{+,+}\right)^*$ for the backward eigen wave, then from boundary conditions we can find

$$G_{1,y}^{\pm,\mp} = -\frac{\alpha_p^{(-)}}{\alpha_p^{(+)}} u^{\pm} G_{1,y}^{\pm,\pm},$$

$$G_{2,y}^{\pm,+} = \frac{1}{2}\rho^{\pm}\exp(-ik_{2,z}D)\left\{\alpha_p^{(\pm)} - \alpha_p^{(\mp)}\frac{\alpha_p^{(+)}}{\alpha_p^{(-)}}u^{\pm}\right\}G_{1,y}^{\pm,\pm}, \qquad (43)$$

$$G_{2,y}^{\pm,-} = \frac{1}{2}\rho^{\pm}\exp(ik_{2,z}D)\left\{\alpha_p^{(\mp)} - \alpha_p^{(\pm)}\frac{\alpha_p^{(+)}}{\alpha_p^{(-)}}u^{\pm}\right\}G_{1,y}^{\pm,\pm},$$

where

$$u^{\pm} = \frac{\left\{\exp(\pm ik_{1,z}d_1) - \rho^{\pm}\exp(\mp ik_{2,z}d_2)\right\}}{\left\{\exp(\mp ik_{1,z}d_1) - \rho^{\pm}\exp(\mp ik_{2,z}d_2)\right\}}. \qquad (44)$$

Using (43), $N_{q,p}^{(z)}$ can be represented as

$$N_{q,p}^{(z)} = 2\frac{\sigma_{p,1}}{Z_0}G_{1,y}^{+,+}G_{1,y}^{-,-}\left(1 - \frac{\alpha_p^{(-)2}}{\alpha_p^{(+)2}}u^+u^-\right). \qquad (45)$$

If $\rho^- = \rho^{+*}$ (propagation band), then $N_{1,p}^{(z)} = N_{2,p}^{(z)}$.

## APPENDIX 2

The Helmholtz equation, which describes scalar wave propagation suited for electromagnetic wave propagating in a dielectric, is the basic of the wave electromagnetic theory.

This differential equation has no analytical solution in the general case, except for a few special cases. This equation can be transformed into different systems of the first order equations. One of such transformation gives a system from which an approximate solution can be easily obtained [16].

Short description of this transformation is given bellow.

The solution of the equation

$$\frac{d^2 E_y}{dz^2} + \frac{\omega^2}{c^2}\varepsilon(z)E_y = 0 \qquad (46)$$

can be represented as the sum of two new functions

$$E_y = E_y^+ + E_y^-. \qquad (47)$$

By introducing two unknown functions $E_y^+$ and $E_y^-$ instead of the one $E_y$, we can impose an additional condition. This additional condition we write in the form

$$\frac{dE_y}{dz} = g^+(z)E_y^+ + g^-(z)E_y^-, \qquad (48)$$

where $g^+(z)$ and $g^-(z)$ ($g^+(z) \neq g^-(t)$) are arbitrary continuous functions having continuous derivatives.

As $g_1(t) \neq g_2(t)$, from (47), (48) and (46) we can find the derivatives of $\dfrac{dE_y^+}{dz}$ and $\dfrac{dE_y^-}{dz}$

$$\left(g^+ - g^-\right)\frac{dE_y^+}{dz} = -E_y^+\left(\frac{dg^+}{dz} + f_0 + g^+g^-\right) - E_y^-\left(\frac{dg^-}{dz} + f_0 + \left(g^-\right)^2\right),$$

$$\left(g^+ - g^-\right)\frac{dE_y^-}{dz} = E_y^+\left(\frac{dg^+}{dz} + f_0 + \left(g^+\right)^2\right) + E_y^-\left(\frac{dg^-}{dz} + f_0 + g^+g^-\right), \qquad (49)$$

where $f_0 = \dfrac{\omega^2}{c^2}\varepsilon$.

If we choose

$$\left(g^{\pm}\right)^2 + f_0 g^{\pm} = 0, \qquad (50)$$

then the system (49) takes the form

$$\frac{dE_y^+}{dz} = \left(i\frac{\omega}{c}\sqrt{\varepsilon(z)} - \frac{1}{4\varepsilon}\frac{d\varepsilon}{dz}\right)E_y^+ + \frac{1}{4\varepsilon}\frac{d\varepsilon}{dz}E_y^-,$$
$$\frac{dE_y^-}{dz} = -\left(i\frac{\omega}{c}\sqrt{\varepsilon(z)} + \frac{1}{4\varepsilon}\frac{d\varepsilon}{dz}\right)E_y^- + \frac{1}{4\varepsilon}\frac{d\varepsilon}{dz}E_y^+. \tag{51}$$

## REFERENCES

1 E.Bahar Electromagnetic wave propagation in inhomogeneous multilayered structures of arbitrary thickness-Full wave solutions. Journal of Mathematical Physics, 1973, 14(8), 1030–1036, http://dx.doi.org/10.1063/1.1666434; Depolarization of electromagnetic waves excited by distributions of electric and magnetic sources in inhomogeneous multilayered structures of arbitrarily varying thickness. Generalized field transforms. Journal of Mathematical Physics, 1973, 14(11), 1502–1509; http://dx.doi.org/10.1063/1.1666216; Depolarization of electromagnetic waves excited by distributions of electric and magnetic sources in inhomogeneous multilayered structures of arbitrarily varying thickness. Full wave solutions. Journal of Mathematical Physics, 1973, 14(11), 1510–1515; http://dx.doi.org/10.1063/1.1666217; Depolarization in nonuniform multilayered structures-full wave solutions. Journal of Mathematical Physics, 1974, 15(2), 202–208; http://dx.doi.org/10.1063/1.1666621
2 P. Lee, Uniform and graded multilayers as x-ray optical elements. Applied Optics, 1983, 22(8), 1241, doi:10.1364/ao.22.001241
3 Pochi Yeh, Optical waves in layered media, John Wiley & Sons, 1988.
4 N.-H. Sun, J.-J. Liau, Y.-W. Kiang, S.-C. Lin, R.-Y. Ro, J.-S. Chiang, H.-W. Chang. Numerical analysis of apodized fiber bragg gratings using coupled mode theory, PIER 99, 289–306, 2009
5 Z. Jakšic, R. Petrovic a , Z. Djuric, Nonuniform photonic crystals for multicolor infrared photodetection, Journal of Optoelectronics and Advanced Materials Vol. 4, No. 1, March 2002, p. 13 – 20
6 S. G.Johnson, P.Bienstman, M. A.Skorobogatiy, M.Ibanescu, E.Lidorikis, J. D. Joannopoulos, Adiabatic theorem and continuous coupled-mode theory for efficient taper transitions in photonic crystals. Physical Review E, 2002, 66, 066608
7 E. H. Khoo, A. Q. Liu, and J. H. Wu Nonuniform photonic crystal taper for high-efficiency mode coupling. Optics Express, Vol. 13, Issue 20, pp. 7748-7759, (2005), https://doi.org/10.1364/OPEX.13.007748
8 Tian, H., Ji, Y., Li, C., & Liu, H. Transmission properties of one-dimensional graded photonic crystals and enlargement of omnidirectional negligible transmission gap. Optics Communications, 2007, 275(1), 83–89
9 Luca Dal Negro (Editor), Optics of aperiodic structures, Taylor & Francis Group, 2014
10 Ory Schnitzer Waves in slowly varying band-gap media, SIAM Journal on Applied Mathematics, 2017, Vol.77, Iss.4, pp.1516-1535, https://doi.org/10.1137/16M110784X
11 M. Maksimovic, M. Hammer, E. van Groesen, Field representation for optical defect resonances in multilayer microcavities using quasi-normal modes, Optics Communications, 2008, Vol. 281, pp. 1401-1411
12 R.M. Feshchenko, A theory of reflective X-ray multilayer structures with graded period and its applications, https://doi.org/10.48550/arXiv.2109.05330, 2021
13 Grigoryan, L. S., Mkrtchyan, A. R., Khachatryan, H. F., Arzumanyan, S. R., & Wagner, W. Radiation from a charged particle-in-flight from a laminated medium to vacuum. Journal of Physics: Conference Series, 2010, 236
14 Zheng Gong, Jialin Chen, Ruoxi Chen, Xingjian Zhu, Chan Wang, Xinyan Zhang, Hao Hu, Yi Yang, Baile Zhang, Hongsheng Chen, Ido Kaminer, Xiao Lin, Interfacial Cherenkov radiation from ultralow-energy electrons, PNAS 2023, Vol. 120, No. 38, e2306601120, https://doi.org/10.1073/pnas.2306601120
15 M.I. Ayzatsky Coupled-mode theory for non-periodic structured waveguides https://doi.org/10.48550/arXiv.2309.06280, 2023
16 M.I. Ayzatsky A note on the transformation of the linear differential equation into a system of the first order equations. https://arxiv.org/abs/1803.03124, 2018